\documentclass[12pt]{article}
\usepackage{graphicx}
\usepackage{amsmath}

\newcommand\pubnumber{Article 44 in eConf C1304143}
\newcommand\pubdate{\today}

\def\msu{Department of Physics and Astronomy, Michigan State University, Michigan, USA.}
\def\unf{Department of Physics, University of North Florida, Florida, USA.}
\def\umd{Department of Physics, University of Maryland, Maryland, USA.}
\def\lab1{Laboratoire Univers et Particules de Montpellier, Universit\'e Montpellier 2, CNRS/IN2P3, CC 72, Place Eug\`ene Bataillon,\\
F-34095 Montpellier Cedex 5, France.}
\def\lanl{Los Alamos National Laboratory, New Mexico, USA.}

\def\nat{\emph{Nature}}

\def\prd{{\em Phys. Rev.} D}

\def\Title#1{\begin{center} {\Large #1 } \end{center}}
\def\Author#1{\begin{center}{ \sc #1} \end{center}}
\def\Address#1{\begin{center}{ \it #1} \end{center}}

\newcommand\pubblock{\rightline{\begin{tabular}{l} \pubnumber\\
         \pubdate  \end{tabular}}}
\newenvironment{Abstract}{\begin{quotation}  }{\end{quotation}}
\newenvironment{Presented}{\begin{quotation} \begin{center}
             PRESENTED AT\end{center}\bigskip
      \begin{center}\begin{large}}{\end{large}\end{center} \end{quotation}}

\def\beq{\begin{equation}}
\def\eeq#1{\label{#1}\end{equation}}
\def\eeqn{\end{equation}}

\def\beqa{\begin{eqnarray}}
\def\eeqa#1{\label{#1}\end{eqnarray}}
\def\eeqan{\end{eqnarray}}

\let\bar=\overbar

\def\Dslash{\not{\hbox{\kern-4pt $D$}}}
\def\dslash{\not{\hbox{\kern-2pt $\del$}}}

\def\msb{{\bar{\ssstyle M \kern -1pt S}}}

\begin{document}
\begin{titlepage}
\pubblock

\vfill
\Title{Milagro Limits on the Rate-Density of Primordial Black Holes}
\vfill
\Author{T. U. Ukwatta, D. Stump, J. T. Linnemann, and K. Tollefson}
\Address{\msu}
\Author{V. Vasileiou}
\Address{\umd}
\Address{\lab1}
\Author{G. Sinnis}
\Address{\lanl}
\Author{J. H. MacGibbon}
\Address{\unf}
\Author{For the Milagro Collaboration.}

\vfill
\begin{Abstract}
Primordial Black Holes (PBHs) created early in the universe are dark matter candidates. One method of detecting these PBHs is through their Hawking radiation. PBHs created with an initial mass of $5.0 \times 10^{14}$ g should be evaporating today with bursts of high-energy particles, including gamma radiation in the GeV - TeV energy range. The Milagro high energy observatory, which operated from 2000 to 2008, is sensitive to the high end of the PBH evaporation gamma ray spectrum. Due to its large field-of-view, more than 90\% duty cycle and sensitivity up to 100 TeV gamma rays, the Milagro observatory is ideally suited for the direct search of PBH bursts. Based on a search in Milagro data, we report PBH upper limits according to the standard model.
\end{Abstract}
\vfill
\begin{Presented}
Huntsville Gamma Ray Burst Symposium\\
Nashville, Tennessee, April 14--18, 2013
\end{Presented}
\vfill
\end{titlepage}
\def\thefootnote{\fnsymbol{footnote}}
\setcounter{footnote}{0}
\section{Introduction}

In the present universe, stellar processes can create black holes (BHs) with masses greater about $3 M_{\odot}$. Processes in the Early Universe, however, may have created BHs with sub-stellar masses. Density fluctuations in the early
universe could have created BHs with arbitrarily small
masses down to the Planck scale~\cite{Carr2010}. These BHs
are referred to as Primordial Black Holes (PBHs).

In a ground-breaking theoretical study, Hawking showed that due to
quantum-gravitational effects, BHs possess a temperature~\cite{Hawking1974}.
In addition, he showed that the BH temperature is inversely proportional
to its mass~\cite{Hawking1974}. The immediate implication of this finding is the
realization that a BH with a temperature higher than its surrounding
environment will evaporate. Stellar mass BHs have temperatures much lower than that of the present Cosmic Microwave Background radiation and they will not lose mass through evaporation. On the other hand, PBHs with initial masses smaller than $\sim \, 5.0 \times 10^{14}$ g
have already evaporated and may contribute to the extragalactic
background radiation. PBHs with an initial mass somewhat greater than $\sim \, 5.0 \times 10^{14}$ g should be evaporating now~\cite{MacGibbon1991a}
with bursts of high-energy particles, including gamma radiation in
the MeV -- TeV energy range, making them candidate gamma ray burst
(GRB) progenitors.

The properties of the final PBH burst depend on the physics governing the
production and decay of high-energy particles. As the BH
evaporates and loses mass over its lifetime, its temperature and the
number of distinct particle species emitted increase. The higher the number
of fundamental particle degrees of freedom, the faster and more powerful
will be the final burst from the PBH, with the spectral details differing
according to the particle physics model. Hence, the nature of the final
burst will also provide information on the correct model of high-energy
particle physics~\cite{Carr2010}.

Various detectors have searched for PBHs events using direct and
indirect methods. These methods probe the PBH distribution on various
distance scales. One can probe the PBH rate density at the
cosmological scale using the 100 MeV extragalactic gamma ray
background, which results in a limit of
$< \, 10^{-6}$ ${\rm pc^{-3} yr^{-1}}$~\cite{Carr2010} assuming
no PBH clustering. On the galactic scale, if PBH are clustered
in the Galaxy, we would expect to see anisotropy in the 100 MeV
gamma ray measurements. Indeed such a anisotropy has been detected
which results in a PBH limit of $<$ 0.42 ${\rm pc^{-3} yr^{-1}}$~\cite{Wright1996}.
On the kiloparsec scale the best limit comes from the
antiproton flux studies which is $<$ 0.0012 ${\rm pc^{-3} yr^{-1}}$~\cite{Abe2012}.
It is important to note that however, the antiproton background limit depends
on the distribution of PBHs within the Galaxy and the propagation of
antiprotons through the Galaxy, as well as the production and the propagation of
the secondary antiproton component which is produced by interactions of cosmic-ray
nuclei with the interstellar gas. On the parsec scale, the limits are set by searches for the direct detection of individual bursting PBHs and are independent of the assumptions of PBH clustering. The best limits come from the Very High Energy (VHE) searches done by the Imaging
Air Cherenkov Telescopes (IACTs) and Extensive Air Shower arrays (EAS). For the parsec scale,
the current best limit is $<$ $4.6 \times 10^{5}$ ${\rm pc^{-3} yr^{-1}}$~\cite{Amenomori1995}.

In this paper, we present new PBH limits based on a search done using
the data from the Milagro observatory. These limits are obtained
assuming standard model of Hawking radiation and particle physics~\cite{MacGibbon1990, Halzen1991}.

\section{Milagro Observatory}

Milagro is a water Cherenkov gamma ray observatory (EAS type) sensitive
to the gamma rays in the energy range $\sim$ 100 GeV to 100 TeV.
This observatory is located near Los Alamos, New Mexico, USA at
latitude $35.9^{\circ}$ north, longitude $106.7^{\circ}$ west and altitude of 2630 m,
and was operational from 2000 to 2008~\cite{Atkins2000}. The
Milagro detector had two components: a central rectangular 60 m $\times$ 80 m $\times$ 7 m
reservoir filled with purified water and an array of 175 smaller outrigger (OR) tanks
surrounding the reservoir. These OR tanks were distributed
over an area of 200 m $\times$ 200 m. The reservoir was light-tight
and instrumented with two layers of 8'' photomultiplier tubes (PMTs).
The top layer consisted of 450 PMTs (air-shower or AS layer) 1.5 m under water
and the bottom layer had 273 PMTs (muon or MU layer) 6 m under the water surface.
Each outrigger tank contained one PMT. The observatory detected VHE gamma rays
by detecting the Cherenkov light produced by the secondary particles from the gamma ray
air shower as they pass through the water. Various components of the
detector were used to measure the direction of the gamma ray photon and
to reduce the background due to hadron-induced showers. The Milagro
detector did not have good energy resolution and the median energy
of the gamma rays detected from a Crab-like source was $\sim$ 3 TeV.
However, because of its large field-of-view of $\sim$2 sr and a high-duty
cycle which was over 90\%, the Milagro observatory is an ideal instrument to search for
emission from PBH candidates.

\section{Methodology} \label{methodology}

\subsection{PBH Spectrum}

The temperature ($T$) of a black hole depends on the remaining lifetime ($\tau$) of the black hole
(the time left until the total evaporation is completed) as follows~\cite{Petkov2010}:
\begin{equation} \label{tempEq}
T = \bigg[4.7 \times 10^{11} \, \bigg(\frac{\rm{1 sec}}{\tau}\bigg) \bigg]^{1/3}\,\,\,\rm GeV.
\end{equation}
For BHs with temperatures greater than several GeVs at the start of the observation,
the time--integrated photon flux can be parameterized (for $E > \sim$ 10 GeV) as~\cite{Petkov2010}
\begin{equation} \label{photonEq}
\frac{dN}{dE} \approx 9 \times 10^{35} \begin{cases}
\big(\frac{1 GeV}{T}\big)^{3/2}\big(\frac{1 GeV}{E}\big)^{3/2},\,\,\,\,E<T \\
\big(\frac{1 GeV}{E}\big)^{3},\,\,\,\,E\ge T
\end{cases}
\end{equation}
where $E$, the gamma ray photon energy, is measured in GeV.

\subsection{Detectable Volume Estimation}

The expected number of photons detectable by an
observatory on the ground from a PBH burst of duration $\tau$ seconds
at a distance $r$ and zenith angle $\theta$ is
\begin{equation} \label{countsEq}
\mu(r, \theta, \tau) = \frac{(1-f)}{4 \pi r^2} \int_{E_1}^{E_2} \,\frac{dN(\tau)}{dE}\, A(E,
\theta)\,dE.
\end{equation}
Here $f$ is the dead time of the detector and $dN(\tau)/dE$ is
the gamma ray emission spectrum integrated over remaining time from $\tau$ to 0.
The values $E_1$ and $E_2$ correspond to the lower and upper bounds
of the energy range searched and $A(E,\theta)$ is the
effective area of the detector as a function of photon energy and
zenith angle. Typically the function $A(E, \theta)$ is obtained
from a simulation of the detector. For Milagro, we have parameterized
the effective area for three zenith angle bands as $A(E, \theta) = 10^{a(\log E)^2 + b \log E + c} \,\,\, \rm m^2$
and parametrization parameters are given in Table~\ref{para_table}.
\begin{table}[h]
\begin{center}
\begin{tabular}{|c|c|c|c|}
\hline Zenith Angle Range ($\theta$) & a & b & c   \\ \hline
0$^{\circ}$ - 15$^{\circ}$  ($\theta_1$) & -0.4933 & 4.7736 & -2.4272 \\ \hline
15$^{\circ}$ - 30$^{\circ}$ ($\theta_2$) & -0.5037 & 5.0102 & -3.4015 \\ \hline
30$^{\circ}$ - 45$^{\circ}$ ($\theta_3$) & -0.4273 & 4.7931 & -4.3030 \\ \hline
%45$^{\circ}$ - 60$^{\circ}$ ($\theta_4$) & -0.1833 & 3.3077 & -3.8224 \\ \hline
\end{tabular}
\caption{Effective area parametrization parameters for various zenith angle bands.}
\label{para_table}
\end{center}
\end{table}

%It is an analytical calculation using the MacGibbon \& Webber model PBH emission model which does not include a PBH chromosphere, hence predicting an unrestricted emission at $>$TeV energies. We first calculated the predicted emission spectrum by an evaporating PBH in the 30GeV-100TeV energy range versus its remaining lifetime (or temperature) and distance. Then, using the response of the Milagro detector, we calculated the predicted number of events corresponding to the incoming gamma-ray flux. Comparing this number with the expected Milagro backgrounds, and taking into account Poisson fluctuations on the actual number of detected events (from the evaporating PBH), we calculated the minimum mean number of detected events required to ensure a detection with 99\% (frequentist) probability.

The minimum number of counts needed for a detection, $\mu_{\circ}(\tau)$, for different burst durations
are estimated by finding the number of counts required over the background for a $5 \sigma$ detection
with 99\% probability after trials correction. This has been done using a Monte Carlo simulation
in the ref~\cite{Vasileiou2008} and we have used those numbers in our calculation
(see Table~\ref{mu_limit_table}).

%\begin{table}[h]
%\begin{center}
%\begin{tabular}{|l|c|c|}
%\hline Burst Duration (s) & $\mu_{\circ}$ \\ \hline
%0.001 & 11  \\ \hline
%0.01 & 16  \\ \hline
%0.1 & 22  \\ \hline
%1.0 & 35  \\ \hline
%10.0 & 65  \\ \hline
%100.0 & 150 \\ \hline
%\end{tabular}
%\caption{Counts needed over the background for a $5\sigma$ detection with 99\% probability
%for various burst durations ($\tau$).}
%\label{mu_table}
%\end{center}
%\end{table}

\begin{table}[h]
\begin{center}
\begin{tabular}{|l|c|c|}
\hline Burst Duration (s) & $\mu_{\circ}$ & $UL_{99}$ ($\rm pc^{-3} yr^{-1}$) \\ \hline
0.001 & 11 & 2.8 $\times 10^5$ \\ \hline
0.01 & 16 & 1.1 $\times 10^5$ \\ \hline
0.1 & 22 & 4.8 $\times 10^4$ \\ \hline
1.0 & 35 &  3.3 $\times 10^4$ \\ \hline
10.0 & 65 &  3.4 $\times 10^4$ \\ \hline
100.0 & 150 &  6.2 $\times 10^4$ \\ \hline
\end{tabular}
\caption{Counts needed over the background, $\mu_{\circ}(\tau)$, for a $5\sigma$ detection with 99\% probability
and calculated 99\% confidence upper limits for various burst durations ($\tau$). }
\label{mu_limit_table}
\end{center}
\end{table}

By substituting $\mu_{\circ}$ values corresponding to various burst durations into
Equation~\ref{countsEq} and solving for $r$, we calculate the maximum distance
from which a PBH burst could be detected by the Milagro observatory for the
three zenith bands and for various burst durations,
\begin{equation} \label{distanceEq}
r_{\rm max}(\theta_i, \tau) = \sqrt{ \frac{(1-f)}{4 \pi \mu_{\circ}(\tau)} \int_{E_1}^{E_2} \,\frac{dN}{dE}\, A(E,
\theta_i)\,dE}.
\end{equation}
Denoting the field-of-view of the  detector by FOV($\theta_{i}$)=$2 \pi (1-\cos \theta_{i,\,\rm max})$ sr, the detectable volume is then
\begin{equation} \label{volueEq1}
V(\tau) = \sum_{i} V(\theta_{ i}, \tau) = \frac{4}{3} \pi \sum_{i} r_{\rm max}^3(\theta_{ i}, \tau) \times \frac{{\rm effFOV}(\theta_{i})}{4\pi}
\end{equation}
where $\theta_{i}$ refers to zenith angle band and $\theta_{i,\, \rm max}$ corresponds to the maximum zenith angle in band $i$. The effFOV
is the effective field-of-view for the given zenith angle band. We calculate this by subtracting the FOV of the smaller band from the larger
band as shown below:
\begin{eqnarray} \label{volueEq2}
V(\tau) & = & \frac{1}{3} \bigg[ r_{\rm max}^3(\theta_1, \tau) \cdot {\rm FOV(\theta_1)} + r_{\rm max}^3(\theta_2, \tau) [{\rm FOV(\theta_2)-FOV(\theta_1)}] \nonumber \\
  &   &  + \, r_{\rm max}^3(\theta_3, \tau) [{\rm FOV(\theta_3)-FOV(\theta_2)}] \bigg]
\end{eqnarray}

%\begin{eqnarray} \label{volueEq2}
%V(\tau) & = & \frac{1}{3} \bigg[ r_{\rm max}^3(\theta_1, \tau) \cdot {\rm FOV(\theta_1)} + r_{\rm max}^3(\theta_2, \tau) [{\rm FOV(\theta_2)-FOV(\theta_1)}] \nonumber \\
%  &   &  + r_{\rm max}^3(\theta_3, \tau) [{\rm FOV(\theta_3)-FOV(\theta_2)}] + r_{\rm max}^3(\theta_4, \tau) [{\rm FOV(\theta_4)-FOV(\theta_3)}] \bigg]
%\end{eqnarray}

%\begin{array}{c}
%V = \frac{1}{3} \bigg[ r_{\rm max}(\theta_1)^3 \cdot {\rm FOV(\theta_1)} + r_{\rm max}(\theta_2)^3 \cdot [{\rm FOV(\theta_2)-FOV(\theta_1)}]\\
% + r_{\rm max}(\theta_3)^3 \cdot [{\rm FOV(\theta_3)-FOV(\theta_2)]}
% + r_{\rm max}(\theta_4)^3 \cdot [{\rm FOV(\theta_4)-FOV(\theta_3)]} \bigg].
%\end{array}

%\begin{equation} \label{volueEq2}
%V = \frac{1}{3} \bigg[ r_{\rm max}(\theta_1)^3 \cdot {\rm FOV(\theta_1)} + r_{\rm max}(\theta_2)^3 \cdot [{\rm FOV(\theta_2)-FOV(\theta_1)}] \\
% + r_{\rm max}(\theta_3)^3 \cdot [{\rm FOV(\theta_3)-FOV(\theta_2)]}
% + r_{\rm max}(\theta_4)^3 \cdot [{\rm FOV(\theta_4)-FOV(\theta_3)]} \bigg].
%\end{equation}

\subsection{Upper Limit Estimation}

If PBHs are uniformly distributed in the solar neighborhood, the
X\% confidence level upper limit ($UL_{X}$) to the rate density of
evaporating PBHs can be estimated as
\begin{equation}\label{ulX}
UL_{X} = \frac{n}{V \times P},
\end{equation}
if, at the X\% confidence level, zero bursts are observed over the search duration $P$. Here $V$ is the effective detectable volume and $n$ is the expected upper limit on the number of PBH evaporations given that zero bursts are observed. Note that $P_{\rm Poisson}(0|n)=1-X \Rightarrow n^0 e^{-n}/0! = 1-X \Rightarrow n =
- \ln(1-X) \Rightarrow n = \ln (1/(1-X))$. Thus for $X=99\%$ we have $n=\ln 100 \approx 4.6$ and the upper limit on the evaporating PBH rate density will be

\begin{equation}\label{ul99}
UL_{99} = \frac{4.6}{V \times P}.
\end{equation}

\begin{figure}\label{pbhlimits}
\centering
\includegraphics[width=120mm]{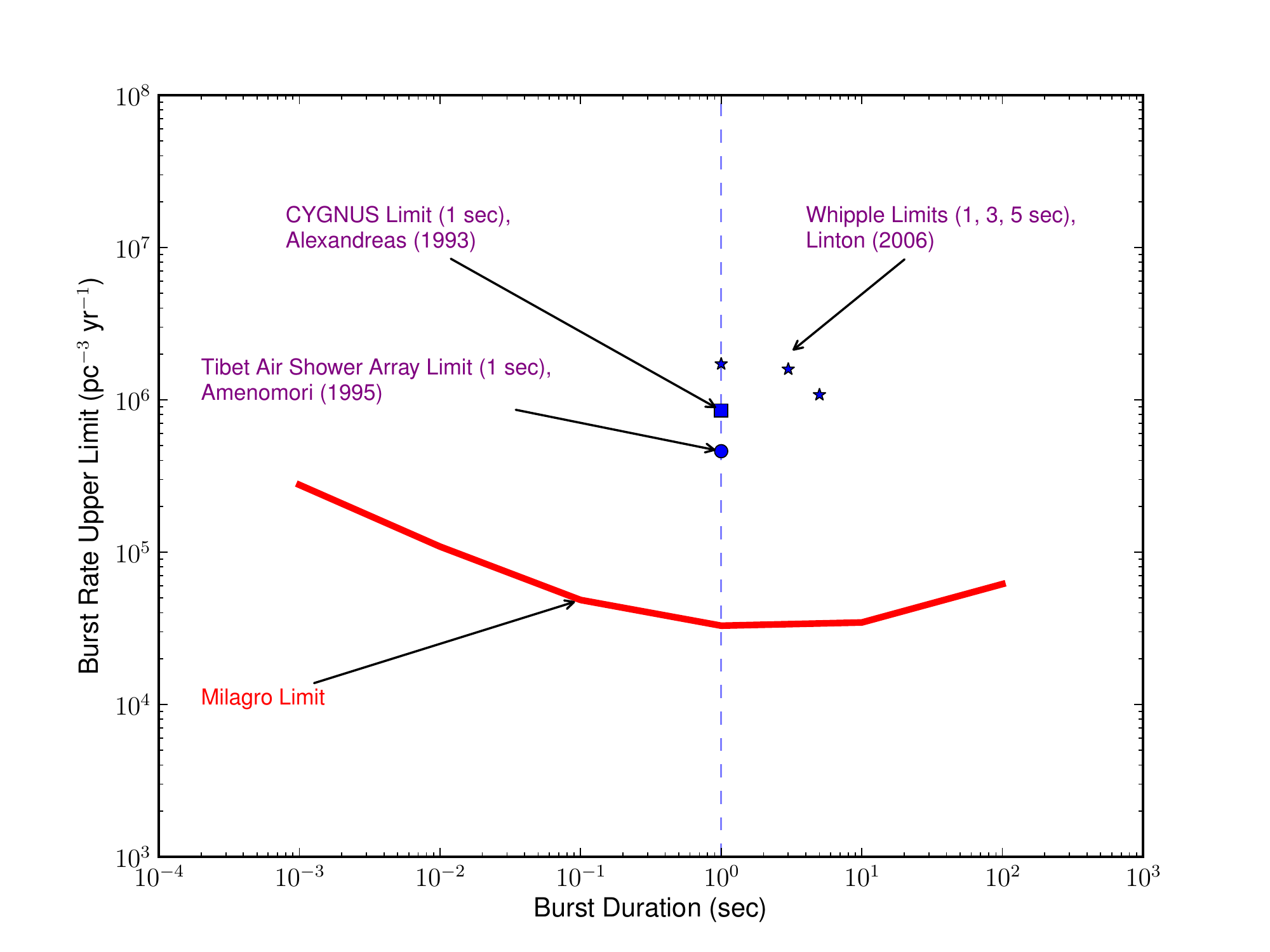}
\caption{PBH Burst Rate Upper Limits from various experiments~\cite{Amenomori1995,Alexandreas1993,Linton2006}.}\label{milagro_eff_aera}
\end{figure}

\section{Results and Discussion}

Even though Milagro had been operating since 2000, for this search only the
last five years of Milagro data have been used: specifically from 03/01/2003 to
03/01/2008. (Due to various detector-related issues some of the data taken
during this period also was not used). The final analysis utilized 1673 days (4.58 years)
worth of good data. This corresponds to $\sim 93$\% of the total Milagro data
collected during the five year period.

Searches were performed for durations ranging from 250 $\mu$s to 6 minutes. No
external triggers were used. The entire reconstructed data set was
systematically searched in time, space and burst duration. No statistically significant
event was observed over the 4.58 years of data. Based on this null detection,
we have calculated upper limits for the PBH rate density using the methodology
outlined in section~\ref{methodology} for various burst durations
(with $E_1$=50 GeV, $E_2$=100 TeV and negligible deadtime). Our results
are shown in Table~\ref{mu_limit_table} and in Figure 1.0.

%According to Figure~\ref{pbhlimits3}, for a null detection, the PBH rate density limit
%from the Milagro observatory is about an order of magnitude better than upper limits
%set by any previous burst searches.

%%%%%%%%%%%%%%%%%%%%%%%%%%%%%%%%%%%%%%%%%%%%%%%%%%%%%%%%%%%%%%%%%%%%%%%%%
%%
%%   use this format to include a LaTeX table  into your paper
%%
%\begin{table}[t]
%\begin{center}
%\begin{tabular}{l|ccc}
%Patient &  Initial level($\mu$g/cc) &  w. Magnet &
%w. Magnet and Sound \\ \hline
% Guglielmo B.  &   0.12     &     0.10      &     0.001  \\
% Ferrando di N. &  0.15     &     0.11      &  $< 0.0005$ \\ \hline
%\end{tabular}
%\caption{Blood cyanide levels for the two patients.}
%\label{tab:blood}
%\end{center}
%\end{table}
%%%%%%%%%%%%%%%%%%%%%%%%%%%%%%%%%%%%%%%%%%%%%%%%%%%%%%%%%%%%%%%%%%%%%%%%%%%

%\Acknowledgements

\end{document}